\pdfoutput=1
\documentclass[10pt,conference]{IEEEtran}
\IEEEoverridecommandlockouts
\usepackage{cite}
\usepackage{amsmath,amssymb,amsfonts}
\usepackage{algorithmic}
\usepackage{graphicx}
\usepackage{textcomp}
\usepackage{xcolor}
\usepackage{tabularx}
\usepackage{authblk}
\usepackage{subfig}
\usepackage[T1]{fontenc}
\usepackage{lmodern}
\usepackage{url}
\usepackage{breqn}
\usepackage[left=0.64in,right=0.64in,bottom=1in,top=0.75in]{geometry}

\def\BibTeX{{\rm B\kern-.05em{\sc i\kern-.025em b}\kern-.08em
    T\kern-.1667em\lower.7ex\hbox{E}\kern-.125emX}}

\newlength\mylength
\newlength\mylengthh
\begin{document}

\title{RAN Cognitive Controller
}

\author[1,2]{Anubhab Banerjee}
\author[1]{Stephen S. Mwanje}
\author[2]{Georg Carle \vspace{-0.28cm}} 
\affil[1]{Nokia Bell Labs, Munich, Germany}
\affil[2]{Dept. of Informatics, Technical University of Munich, Germany}
\affil[ ]{Email:\textit{anubhab.banerjee@tum.de, stephen.mwanje@nokia-bell-labs.com, carle@net.in.tum.de} \vspace{-0.3cm}}

\maketitle

\begin{abstract}
Cognitive Autonomous Networks (CAN) deploys learning based Cognitive Functions (CF) instead of conventional rule-based SON Functions (SF) as Network Automation Functions (NAF) to increase the system autonomy.
These CFs work in parallel sharing the same resources which give rise to conflicts among them which cannot be resolved using conventional rule based approach.
Our main target is to design a Controller which can resolve any type of conflicts among the CFs in a dynamic way. 
\end{abstract}

\begin{IEEEkeywords}
Network management and automation
\vspace{-1.4em}
\end{IEEEkeywords}

\section{Problem Description}
\label{sec:prob}

A typical network is characterized by many control parameters and Key Performance Indicators (KPI). 
For example, a gNB has several control parameters like Transmission Power (TXP), Antenna Tilt (RET) or Antenna Gain, and, multiple observable Key Performance Indicators (KPIs) like downlink throughput, Radio Link Failures (RLF). 
In CAN, each CF focuses on optimizing of one or more KPIs as its objective. 
The CF continuously observes and learns the variation of its objective w.r.t. changes in the control parameters or the external environments, and based on learning, it determines the values of control parameters for which its
objective is optimal in a certain state.

\begin{figure}[!t]
	\centering
	\includegraphics[scale=0.4]{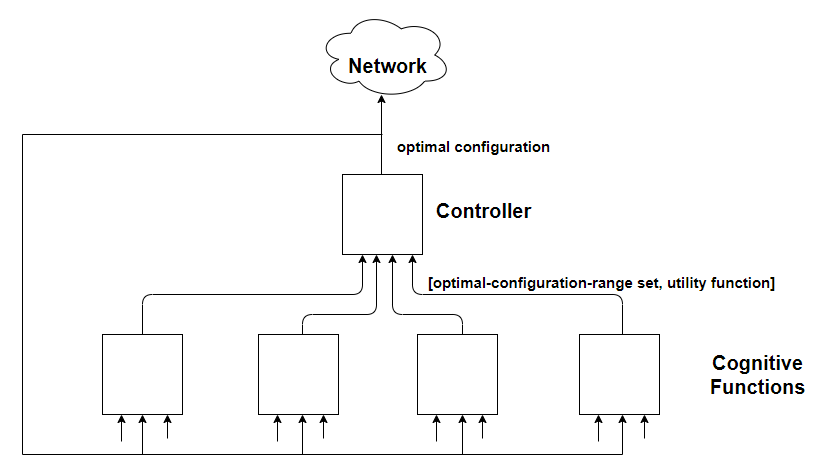}
	\caption{CAN architecture}
	\label{img:can-arch}
\end{figure}

In 5G environment can change very rapidly, and the CFs may detect frequent changes in their respective optimal control parameter values. 
If each CF changes the parameters according to its own will, stability of the system might be compromised. 
This is why we assume, for sake of stability issue, that only the Controller changes the control parameters.
Without going into details, we can abstract a CAN as shown in Fig.~\ref{img:can-arch} in a hierarchical overview where Controller stays one levels above the CFs and makes necessary changes in the control parameters \cite{spects}.

When a control parameter is shared among multiple CFs, and when each CF wants to change the value of the parameter to different degrees of extent, conflicts among them arise.
CFs exhibit three types of conflicts among them - (i) Configuration conflict (while sharing a control parameter or KPI simultaneously), (ii) Measurement conflict (while influencing other CF's output by own action) and (iii) Characteristic conflict (logical dependency) \cite{banerjee2020on}.
As these conflicts cannot be resolved using conventional rule based approach, some new approach is required.
Our goal is to design a Controller which can resolve any type of conflicts among these learning agents and determine the optimal configuration.

\vspace{-1em}

\section{Proposed Solution}
\label{sec:solution}

\begin{figure}[!t]
	\centering
	\includegraphics[scale=0.4]{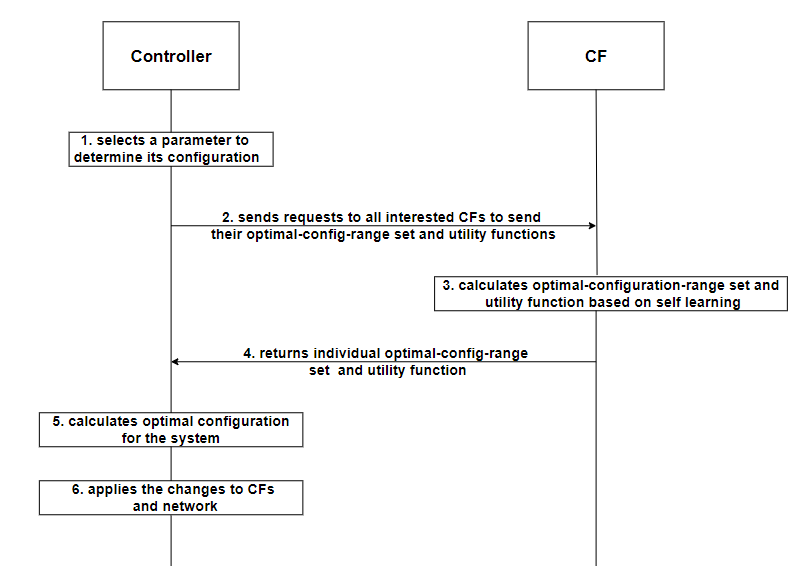}
	\caption{End-to-end workflow}
	\vspace{-1.4em}
	\label{img:e2e}
\end{figure}

We develop a Controller with the following properties -
\begin{itemize}
	\item It can be used to determine the optimal values for all configurations in CAN.
	\item It calculates configuration of the system in a way which optimal for the combined interest of all the CFs.
	\item Its calculation mechanism is very generic and dynamic to be used in real life.
\end{itemize}
In this Section we give a brief overview on how the CFs with the Controller work in CAN and how they exchange necessary information.

A CF always sends two pieces of information to the Controller whenever required - (i) optimal-config-range set and (ii) utility function.
As each CF is a learning agent itself, it can generate its most favorable configuration set based on its learning. 
This set is called optimal-config-range set with the structure: [min\_value, max\_value].
For any parameter value which lies in between min\_value and max\_value, the objective of the CF always lies within a certain percentage of its maximum objective value.
Also, In CAN, different CFs have different objectives with different dimensions. 
For the Controller to understand and compare between different CFs' outputs, it is recommended to convert the outputs in some identical predefined scale. 
Example of such a scale is [0:10], where 0 means the lowest and 10 means the highest achievable value. 
To convert its output to this scale, each CF generates a function called utility function.

A CF is trained with some real life or simulator generated dataset before it becomes operational so that it knows its optimal-config-range set and utility function beforehand.
The system starts with some preloaded configuration determined by the Mobile Network Operator (MNO) based on previous experience.
After the system becomes operational, each CF starts observing its output and begins learning from it. 
After every certain time interval, the CF calculates the optimal-config-range set for a configuration and compares it with the optimal-config-range set calculated in the last cycle. 
If these two sets are identical, the CF continues its learning until the next cycle. 
Otherwise, if there is a change between these two sets, it sends a request to the Controller to recalculate that configuration.
After the Controller receives a request for configuration recalculation from a CF, it sends a message to all CFs in the system, asking for the necessary information.
After the Controller receives all the information, it calculates the optimal configuration as described in the next paragraph and makes necessary changes in the network.
This whole end-to-end workflow is depicted in Fig.~\ref{img:e2e}.

The optimal value is calculated using Nash Social Welfare Function (NSWF) \cite{nswf} which is defined as the product of the individual agent utilities for a particular resource allocation.
The reason behind using NSWF in a resource allocation problem is the solution provided by NSWF balances efficiency and fairness, and it is sensitive to change in overall welfare \cite{anunash}. 
The fairness in allocation can be observed from the lowest difference in utilities obtained.



\section{Implementation and Conclusion}
\label{sec:implement}

\begin{figure}[!t]
	\centering
	\includegraphics[scale=0.46]{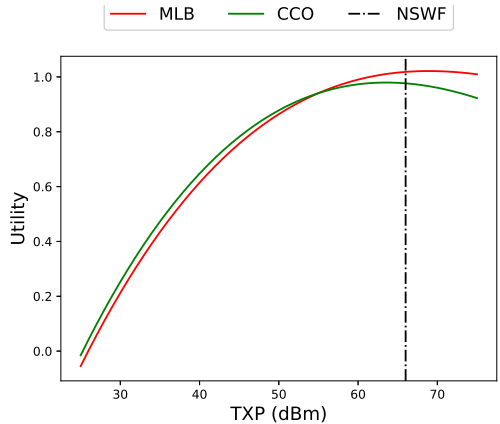}
	\caption{Utility vs TXP}
	\vspace{-1.4em}
	\label{img:txp}
\end{figure}

In this section we discuss how the optimal configurations are calculated in a real life scenario based on a system level simulator called SEASON II which has already been used extensively in previous research works like \cite{tyala1}.
The environment of the simulations is an authentic recreation of a small part of the city of Hamburg with realistic radio propagation models (WINNER+).
In an area of 4 sq. kilometer we deploy 5 cells.
We place 100 static users across all cells randomly for 20 times and collects data.

We assume a CAN deployed at the central cell with two CFs - Mobile Load Balancing (MLB) and Capacity and Coverage Optimization (CCO). 
MLB tries to reduce the load in the cell and CCO tries to maximize the
coverage and capacity of the cell.
Both of them use TXP as their one of the input parameters and have a conflict over it, so in this paper we show how the conflict is resolved and optimal TXP is determined.
We use two neural networks (NN) to implement these two CFs. 
Each CF has 5 fully connected layers, and each layer has 50 nodes with MSE as the loss function and Adaptive learning rate optimizer. 
We collect data from 20 different scenarios and use them for training the NNs,so that after the training is done, each CF can predict the output corresponding to a set of input configurations.
For each CF we use a linear utility scale - for an output value $v$, the corresponding utility value is $\frac{v}{maxv}$ where $maxv$ is the maximum obtained output.

From Fig.~\ref{img:txp} we see that utilities of MLB and CCO are maximum for different TXP values. 
However, when TXP = 66 dBm, their respective utility values are 0.998 and 0.986, and the product of their utilities is maximum.
So, according to NSWF, this value is the optimal TXP for the combined interest of both MLB and CCO i.e., none of them have their maximum utility at this TXP value, but it is a good balance for two of them.
However, problem with this approach is it always assumes the shared parameter has equal impact on both the CFs, which may not be always the case in reality.
As a future direction from here, we plan to incorporate the impact of each parameter on CF so that the optimal solution can be determined more precisly.

\bibliographystyle{unsrt}
\bibliography{references}

\end{document}